\documentclass[aps,prl,twocolumn,superscriptaddress]{revtex4-1}
\usepackage{graphicx}
\newcommand{\wdth}{\gamma} % FWHM in Hz
\setcitestyle{super}

\begin{document}

% Use the \preprint command to place your local institutional report
% number in the upper righthand corner of the title page in preprint mode.
% Multiple \preprint commands are allowed.
% Use the 'preprintnumbers' class option to override journal defaults
% to display numbers if necessary
%\preprint{}

%Title of paper
\title{Anomalous Damping of a Micro-electro-mechanical Oscillator in Superfluid $^3$He-B}

% repeat the \author .. \affiliation  etc. as needed
% \email, \thanks, \homepage, \altaffiliation all apply to the current
% author. Explanatory text should go in the []'s, actual e-mail
% address or url should go in the {}'s for \email and \homepage.
% Please use the appropriate macro foreach each type of information

% \affiliation command applies to all authors since the last
% \affiliation command. The \affiliation command should follow the
% other information
% \affiliation can be followed by \email, \homepage, \thanks as well.
\author{P. Zheng}
%\email[]{Your e-mail address}
%\homepage[]{Your web page}
%\thanks{}
%\altaffiliation{}
\affiliation{Department of Physics, University of Florida, Gainesville, Florida 32611-8440, USA}
\author{W.G. Jiang}
\affiliation{Department of Physics, University of Florida, Gainesville, Florida 32611-8440, USA}
\author{C.S. Barquist}
\affiliation{Department of Physics, University of Florida, Gainesville, Florida 32611-8440, USA}
\author{Y. Lee}
\email[]{yoonslee@phys.ufl.edu}
\affiliation{Department of Physics, University of Florida, Gainesville, Florida 32611-8440, USA}
\author{H.B. Chan}
\affiliation{Department of Physics, The Hong Kong University of Science and Technology, Clear Water Bay, Hong Kong}

%Collaboration name if desired (requires use of superscriptaddress
%option in \documentclass). \noaffiliation is required (may also be
%used with the \author command).
%\collaboration can be followed by \email, \homepage, \thanks as well.
%\collaboration{}
%\noaffiliation

\date{\today}

\begin{abstract}
% insert abstract here
The mechanical resonance properties of a micro-electro-mechanical oscillator with a gap of 1.25~$\mu$m was studied in superfluid $^3$He-B at various pressures. The oscillator was driven in the linear damping regime where the damping coefficient is independent of the oscillator velocity. The quality factor of the oscillator remains low ($Q\approx 80$) down to 0.1\,$T_c$, 4 orders of magnitude less than the intrinsic quality factor measured in vacuum at 4~K. In addition to the Boltzmann temperature dependent contribution to the damping, a damping proportional to temperature was found to dominate at low temperatures. We propose a multiple scattering mechanism of the surface Andreev bound states to be a possible cause for the anomalous damping.
\end{abstract}

% insert suggested PACS numbers in braces on next line
% insert suggested keywords - APS authors don't need to do this
%\keywords{}

%\maketitle must follow title, authors, abstract, \pacs, and \keywords
\maketitle

Over several decades, different families of unconventional superconductors have been discovered. Many of these possess high transition temperatures, which generates much interest from the community pursuing room temperature superconductors. However, the complete microscopic understanding of them still remains a challenge \cite{Norman2011Science1}. Superfluid $^3$He with p-wave spin-triplet pairing is a prime model system to study the unconventional nature of Cooper pairs because the symmetry of the condensate is clearly identified and the properties of the intrinsically pure bulk system are well understood to a quantitative level \cite{Vollhardt1990book}. The early theoretical works \cite{Abrikosov1961JETP1, Larkin1965JETPL1} have revealed the extreme fragility of Cooper pairs against any type of impurity scattering in unconventional superconductors. Interfaces and surfaces also serve as effective pair-breaking agents in these systems, which results in many intriguing surface properties \cite{Buchholtz1981PRB1, Buchholtz1991PRB1}. The surface scattering in unconventional superfluids/superconductors induces quasiparticle mid-gap bound states spatially localized near the surface within the coherence length, $\xi_0$, often called surface Andreev bound states (SABS), accompanying selective suppression of the order parameter components \cite{Ambegaokar1974PRA1, Zhang1987PRB1, Nagato1998JLTP1, Vorontsov2003PRB1}. The detailed structure of SABS has been theoretically investigated for various boundary conditions \cite{Nagato1998JLTP1, Vorontsov2003PRB1}.

In superconductors, tunneling spectroscopy has proven to be a powerful tool for studying the pairing symmetry and surface states \cite{Kashiwaya2000RPP1}. However, the detection of SABS in superfluid $^3$He has been difficult due to the lack of an appropriate probe for the uncharged fluid. Nevertheless, various works have suggested the existence of SABS \cite{Aoki2005PRL1, Nagai2008JPSJ1, Murakawa2009PRL1, Choi2006PRL1, Bradley2016NP1, Castelijns1986PRL1, Volovik2009JETPL1, Davis2008PRL1, Elbs2007JLTP1}. Measurements of transverse acoustic impedance using quartz transducers have been used to investigate SABS \cite{Aoki2005PRL1}. The measured transverse acoustic impedance agrees with theoretical calculations, which provides indirect confirmation of SABS \cite{Nagai2008JPSJ1, Murakawa2009PRL1}. The high resolution heat capacity measurement of $^3$He in a silver heat exchanger was able to identify the contribution from the SABS near the silver surface \cite{Choi2006PRL1}. In the recent experiment by the Lancaster group \cite{Bradley2016NP1}, they linked the absence of the critical velocity of a wire moving without acceleration to the presence of SABS. Recent theoretical studies provide a fresh insight into the nature of SABS \cite{Chung2009PRL1, Nagato2009JPSJ1}. They suggest the anisotropic magnetic response of the film or surface of $^3$He-B with specular boundaries as a direct indicator of Majorana fermions in surface bound states.

Various resonators in direct contact with liquid $^3$He, such as torsional oscillators \cite{Casey2004PRL1}, vibrating wires \cite{Carless1983JLTP1, Guenault1986JLTP1}, tuning forks \cite{Blaauwgeers2007JLTP1, Bradley2009JLTP1}, and moving wires \cite{Bradley2016NP1}, have been successfully utilized to investigate the properties of its normal and superfluid phases. A new direction in the development of the mechanical probes is based on the nanolithography technology, such as micro- and nano-electro-mechanical system (MEMS and NEMS) devices \cite{Gonzalez2013RSI1, Defoort2016JLTP1}. We have developed MEMS devices to study superfluid $^3$He films \cite{Gonzalez2011JLTP1, Zheng2014JPCS1}. Theses devices have also been successfully exploited to study the viscosity of normal liquid $^3$He below 800~mK \cite{Gonzalez2016PRB1}. In this paper, we report the measurement of the damping of a MEMS device in superfluid $^3$He-B which exhibits anomalous low temperature behavior. A plausible physical mechanism involving SABS is conjectured to be responsible for the observed behavior.

\begin{figure}
% Fig. 1
\includegraphics[width=0.76\linewidth]{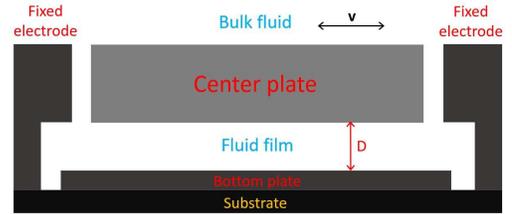}
\caption{A schematic side-view of the MEMS device. A mobile center plate is suspended above the bottom plate by springs (not shown). The gap $D$ between the mobile plate and the bottom plate is 1.25~$\mu$m. The thickness of each layer is shown to scale. The horizontal arrow represents the direction of the oscillation of the shear mode. \label{MEMSimage}}
\end{figure}
The MEMS device used in this measurement has a mobile plate with 2~$\mu$m thickness and 200~$\mu$m lateral size. The plate is suspended above the substrate by four serpentine springs, maintaining a gap of 1.25~$\mu$m. A schematic side-view of the device is shown in Fig.\,\ref{MEMSimage}. When the device is submerged in the fluid, a film is formed between the mobile plate and the substrate, while the bulk fluid is in direct contact with the top surface of the plate. Its in-plane oscillation, called the shear mode, can be actuated and detected by the comb electrodes fabricated on either side of the plate. The details of the devices and the measurement scheme can be found elsewhere \cite{Gonzalez2013RSI1, Barquist2014JPCS1, Zheng2016JLTP1}.

\begin{figure}
% Fig. 2
\includegraphics[width=0.8\linewidth]{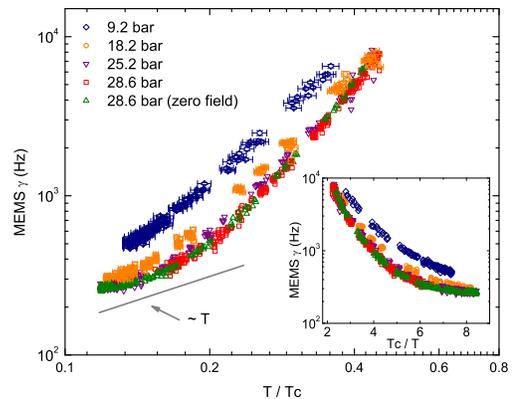}
\caption{(\textit{Color online}) The FWHM of the MEMS as a function of the reduced temperature at various pressures in a log-log scale. For clarity, the error bars are only shown for the data of 9.2 and 18.2~bar. The straight line corresponds to a linear temperature dependence. (\textit{Inset}) The same data in an Arrhenius scale. A straight line in this scale represents a Boltzmann dependence. \label{d1MvsrT}}
\end{figure}

The MEMS device was studied in liquid $^3$He at pressures of 9.2, 18.2, 25.2, and 28.6~bars and cooled down to a base temperature of about 250~$\mu$K by a dilution refrigerator and a copper demagnetization stage. The resonance spectrum of the shear mode was obtained continuously upon warming from the base temperature with a typical warming rate of 30~$\mu$K/hr. The temperature was determined by calibrated tuning fork thermometers \cite{Blaauwgeers2007JLTP1, Bradley2009JLTP1} below 0.6~mK and by a $^3$He melting curve thermometer above. The PLTS-2000 was adopted as the temperature scale \cite{Rusby2002JLTP1}. The uncertainty of temperature measured by the tuning forks is mainly from the calibration process and is represented by error bars in Fig.\,\ref{d1MvsrT}. A magnetic field of 14~mT was applied in the direction perpendicular to the plane of the film except for one of the 28.6~bar measurements. The full width at half maximum (FWHM), $\wdth$, and the resonance frequency, $f_0$, were obtained by fitting the spectrum to the Lorentzian:
\begin{equation}
x=A\frac{\wdth f_0}{\sqrt{(f_0^2-f^2)^2+(\wdth f)^2}},
\end{equation}
where $x$ is the vibration amplitude of the plate, $A=F_0/4\pi^2mf_0\wdth$ is the amplitude of the Lorentzian peak, $F_0$ is the amplitude of the driving force applied on the plate, $m$ is the effective mass of the plate, and $f$ is the frequency of the driving force. The FWHM is proportional to the damping coefficient in the equation of motion of a damped driven harmonic oscillator. The uncertainty from fitting is represented by error bars in Fig.\,\ref{d1MvsrT}. The resonance feature of the MEMS device is sensitive to the temperature \footnote{See Supplemental Material at [URL] for the spectrum at various temperatures.}. For instance, at 28.6~bar its quality factor reaches around 80 when the liquid is cooled to 300~$\mu$K, and decreases rapidly to order of unity near the A-B transition.

The mean free path, $\ell$, of the $^3$He quasiparticles is of the order of 10~$\mu$m at the transition temperature and increases exponentially when the temperature approaches zero due to the isotropic energy gap of $^3$He-B \cite{Vollhardt1990book}. For $T \lesssim 0.4T_c$, $\ell$ becomes larger than any length scale of the MEMS devices, and the MEMS-superfluid system transitions into the ballistic regime. This aspect is verified by the temperature independent resonance frequency observed in this temperature range. At low velocities, the damping has a temperature dependence solely from the density of the quasiparticles which decreases rapidly with temperature as $\exp(-\Delta/k_{_B}T)$ \cite{Guenault1986JLTP1}. Below 0.4\,$T_c$, the energy gap $\Delta$ develops fully to the zero-temperature value, $\Delta_0$. The FWHM of the MEMS is expected to follow
\begin{equation}
\wdth=B\exp(-\Delta_0/k_{_B}T),
\label{Boltzmann}
\end{equation}
where $B$ is the damping amplitude determined by the geometry of the device and the properties of the fluid. 

The intrinsic FWHM measured in vacuum at 4~K, which is 0.071~Hz, is subtracted from the fitted FWHM to yield the FWHM due to the fluid only. This FWHM is plotted as a function of the reduced temperature at various pressures in Fig.\,\ref{d1MvsrT}. At the lowest attainable temperature for 28.6~bar, 280~$\mu$K, the FWHM is around 270~Hz, which is 4 orders of magnitude larger than the intrinsic FWHM and 2 orders of magnitude larger than the TF FWHM in the same condition \footnote{See Supplemental Material at [URL] for the plot of the MEMS $\wdth$ against the TF $\wdth$.}. In contrast, the FWHM of MEMS in superfluid $^4$He below 200~mK is weakly temperature dependent and approaches the intrinsic value \cite{Gonzalez2013JLTP1}. Therefore, the anomalously large damping observed in $^3$He-B is believed to stem from some mechanism other than the scattering of thermal quasiparticles from the bulk. As shown in the inset of Fig.\,\ref{d1MvsrT}, the FWHM does not follow Eqn.\,(\ref{Boltzmann}). Furthermore, the FWHM becomes linear in temperature below $\sim 0.15$\,$T_c$ for the three highest pressures. For 9.2~bar, the linear dependence is not fully developed, probably because of the relatively low $T_c$ at this pressure. This linear temperature dependent term emerging at low temperatures keeps the FWHM of the MEMS from decreasing exponentially as expected. The coefficient of the linear temperature dependent contribution can be extracted from the ratio of FWHM to temperature in the low temperature limit \footnote{See Supplemental Material at [URL] for the plot of $\wdth/T$ against $T$. The Supplemental Material also includes Ref. \cite{Ali2011JLTP1}.}.

\begin{figure}
% Fig. 3
\includegraphics[width=0.8\linewidth]{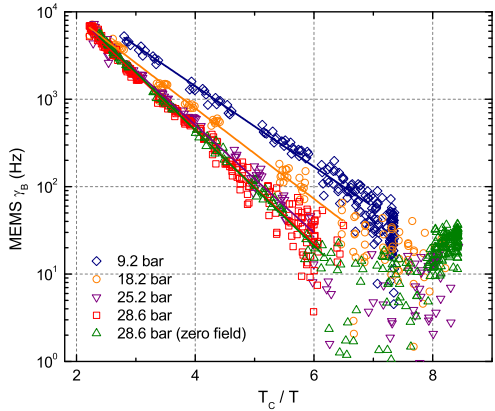}
\caption{(\textit{Color online}) The FWHM of the Boltzmann damping, $\wdth_{_B}$, against the reduced temperature for various pressures in an Arrhenius scale. A linear fit (\textit{straight lines}) in this scale gives the measured energy gap, $\Delta_m$. \label{d2MvsrT}}
\end{figure}
The acquired linear term is then subtracted from the total FWHM. The residual FWHM is plotted against the temperature in an Arrhenius scale in Fig.\,\ref{d2MvsrT}. The linear behavior of the three highest pressures demonstrates a Boltzmann exponential temperature dependence following Eqn.\,(\ref{Boltzmann}) and justifies the assumption that the damping in addition to the expected thermal quasiparticles in the bulk is linear. Since its temperature dependence is not fully developed at low temperatures for 9.2~bar, the linear term is estimated by requiring the residual FWHM to obey Eqn.\,(\ref{Boltzmann}). It was found that the residual FWHM data are sensitive to the choice of the linear coefficient. A 5\% variation in the value is sufficient to skew the dependence of the residual FWHM. Therefore, the total FWHM can be expressed by
\begin{equation}
\wdth=\wdth_{_A}+\wdth_{_B}=\alpha T+B\exp(-\Delta_0/k_{_B}T),
\label{FWHM}
\end{equation}
where $\wdth_{_A}$ is the linear temperature dependent term that dominates at low temperatures, and $\wdth_{_B}$ is the Boltzmann exponential temperature dependent term due to the thermal quasiparticles in the bulk region. Hereafter, the exponential term is called the \textit{Boltzmann} damping and the linear term the \textit{additional} damping.

\begin{figure}
% Fig. 4
\includegraphics[width=0.65\linewidth]{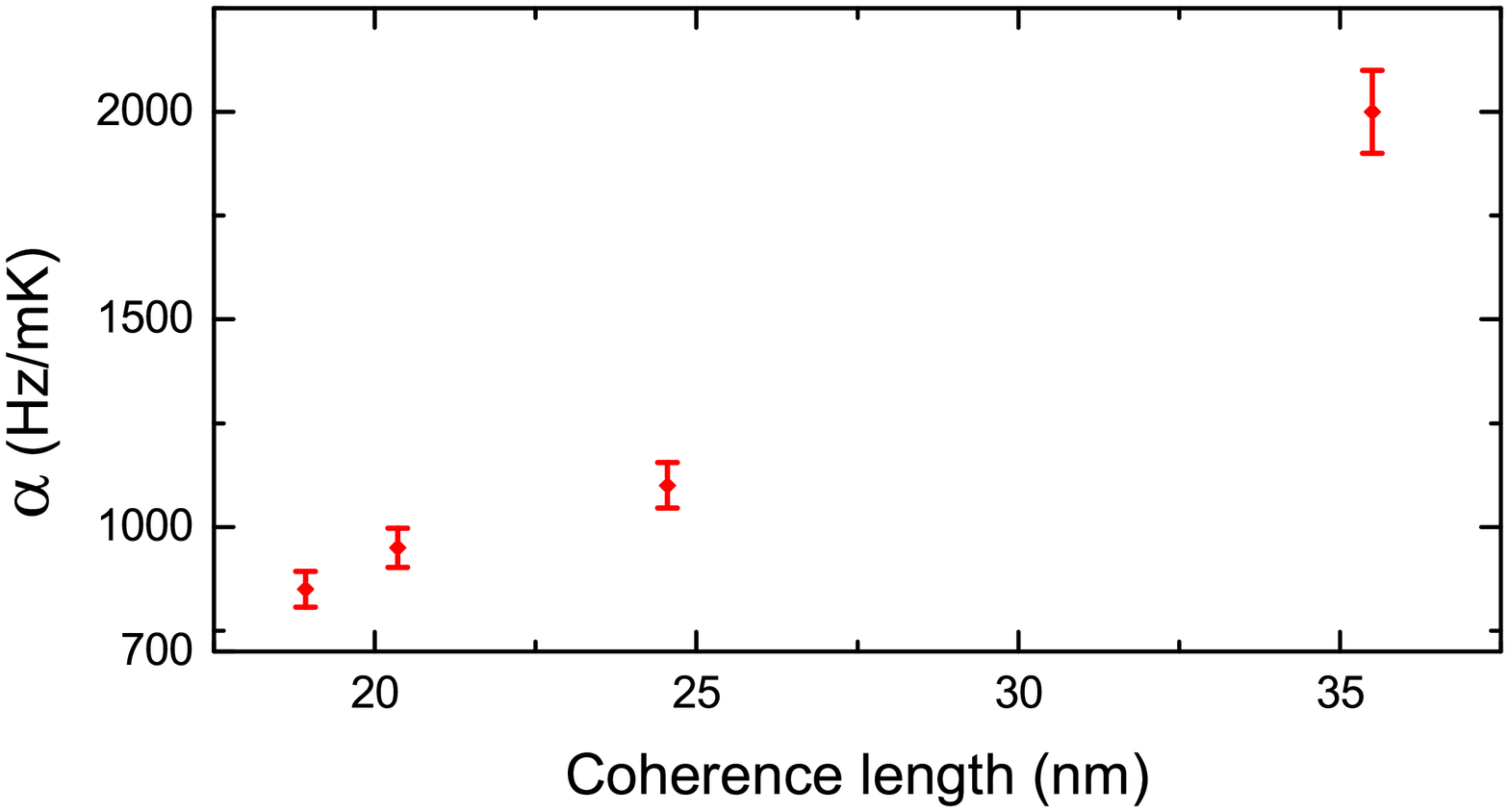}
\includegraphics[width=0.65\linewidth]{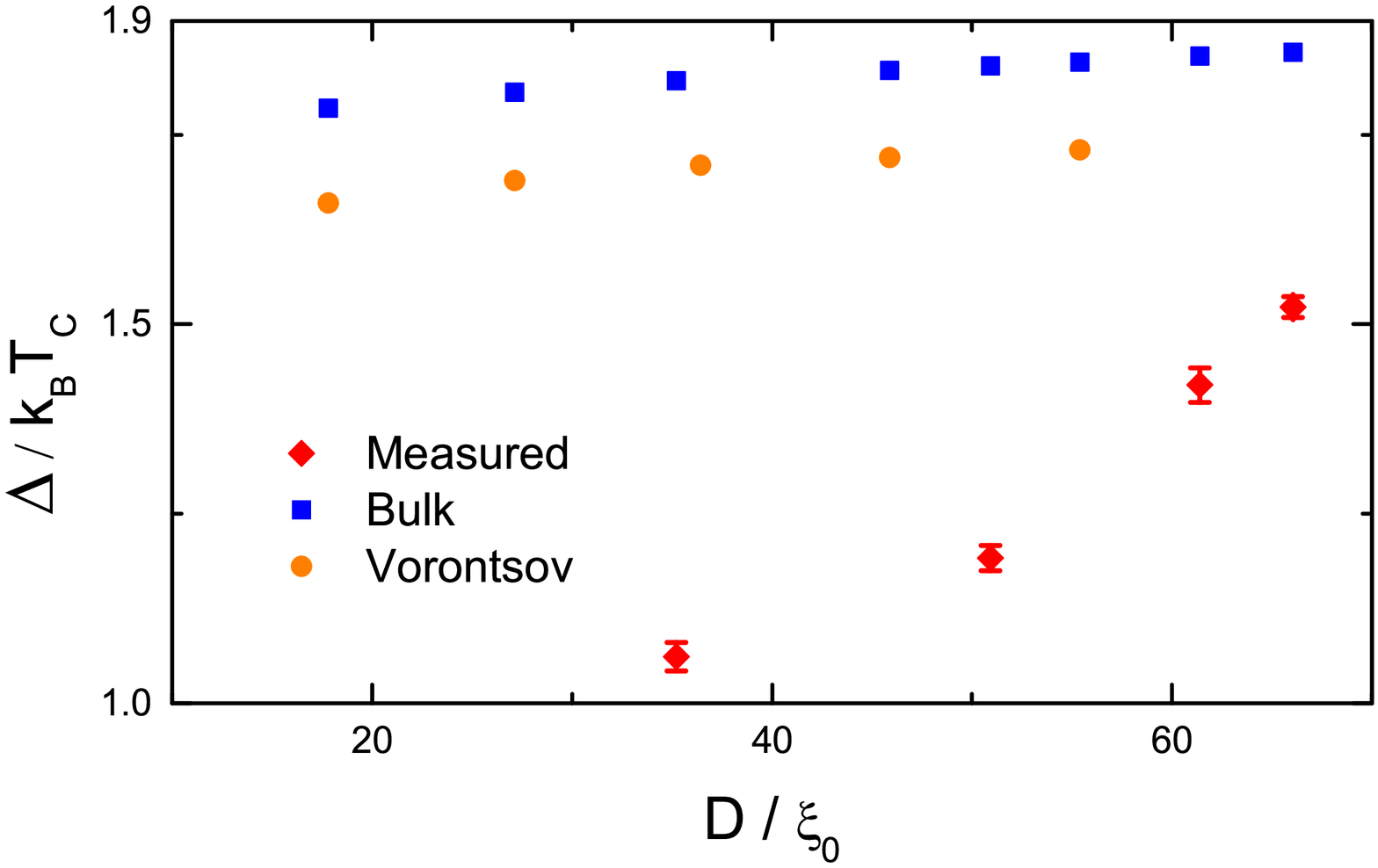}
\caption{(\textit{Top}) The coefficient of the linear temperature term, $\alpha$, against the coherence length of the bulk superfluid. (\textit{Bottom}) The measured energy gap, $\Delta_m$, against the scaled film thickness, $D/\xi_0$. The error bars associated with $\Delta_m$ are from the linear fitting in Fig.\,\ref{d2MvsrT}. Also plotted are the bulk energy gap and the average energy gap in a film calculated by Vorontsov. The bulk gap used here is from the weak coupling plus model \cite{Halperin1990book}. \label{LinCovsCL}}
\end{figure}
The coefficient of the additional damping, $\alpha$, decreases by a factor of two as the pressure increases from 9.2~bar to 28.6~bar (Fig.\,\ref{LinCovsCL}). The linear coefficient seems to have a linear dependence on the coherence length. For the Boltzmann damping, the data in Fig.\,\ref{d2MvsrT} can be fitted to straight lines according to Eqn.\,(\ref{Boltzmann}) to get $\Delta_m$, the measured energy gap. It was found that $\Delta_m$ is much less than the known bulk value at the corresponding pressure. The pressure dependence of $\Delta_m$ is presented in terms of the BCS coherence length $\xi_0=\hbar v_{F}/\pi\Delta_0$, where $v_{_F}$ is the Fermi velocity (Fig.\,\ref{LinCovsCL}). The measured energy gap is suppressed from the bulk value and shows a strong pressure dependence. It decreases monotonically with the scaled film thickness, $D/\xi_0$, since a larger effective thickness gives more space for the order parameter to recover to its bulk value, hence a larger overall energy gap for the film. Also shown in the plot is a calculation which evaluates the energy gap by averaging the parallel and perpendicular components of the order parameter for a superfluid film with both boundaries diffusive \footnote{Private communication with A. B. Vorontsov.}. Our measurement shows a much stronger suppression and pressure dependence of the energy gap than the theoretical estimation.

One might argue the local heating from the MEMS device is responsible for the additional damping. However, the film formed in our MEMS device is in good thermal contact with the surrounding bulk because of the open geometry. Furthermore, the temperature rise due to the heat dissipation is negligible. For instance, at 0.2\,$T_c$ and 9.2~bar, when the center plate is oscillating at a velocity of 1.4~mm/s, the damping force on the plate is about 2~nN, which results in a dissipation power of about 3~pW. In addition, all the measurements were performed in the linear regime where the damping coefficient was independent of the excitation \footnote{See Supplemental Material at [URL] for the linear velocity-force curve.}. Any heating effect would have resulted in the increase of the FWHM at higher excitations.

It is also unlikely that the additional damping comes from the vortices around the MEMS devices or other topological objects as suggested by Winkelmann \textit{et al}.\ \cite{Winkelmann2006PRL1}, because multiple independent cooldowns produced consistent spectra at a given temperature and pressure. During each thermal cycle the MEMS device was driven at high velocities beyond the linear regime where heating effect was clearly observed. The severe heating and the high velocity should have altered vortex lines or other topological objects around the device. But after a reasonable relaxation time, the spectrum always recovers to the shape right before the heating.

\begin{figure}
% Fig. 5
\includegraphics[width=0.50\linewidth]{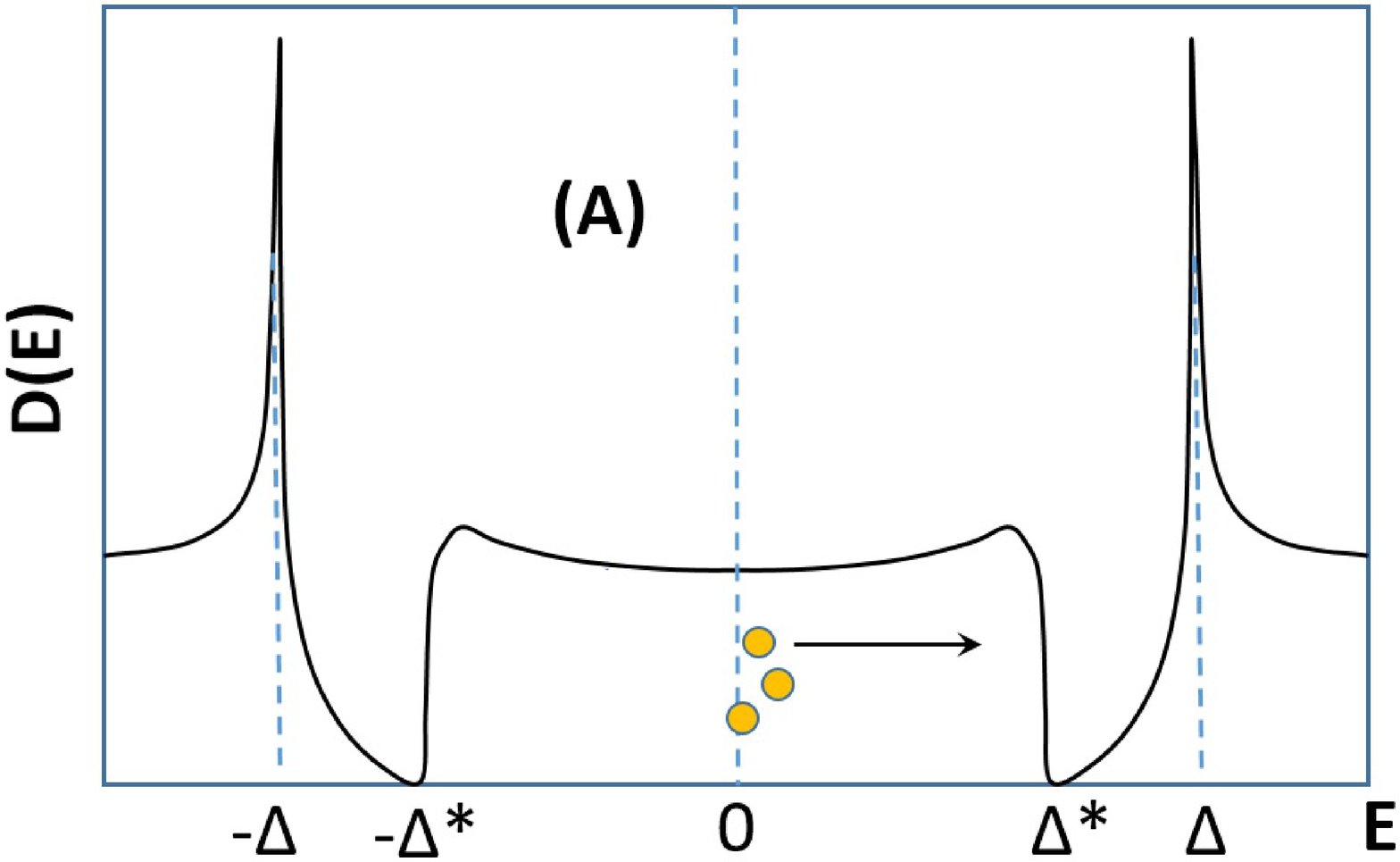}
\includegraphics[width=0.80\linewidth]{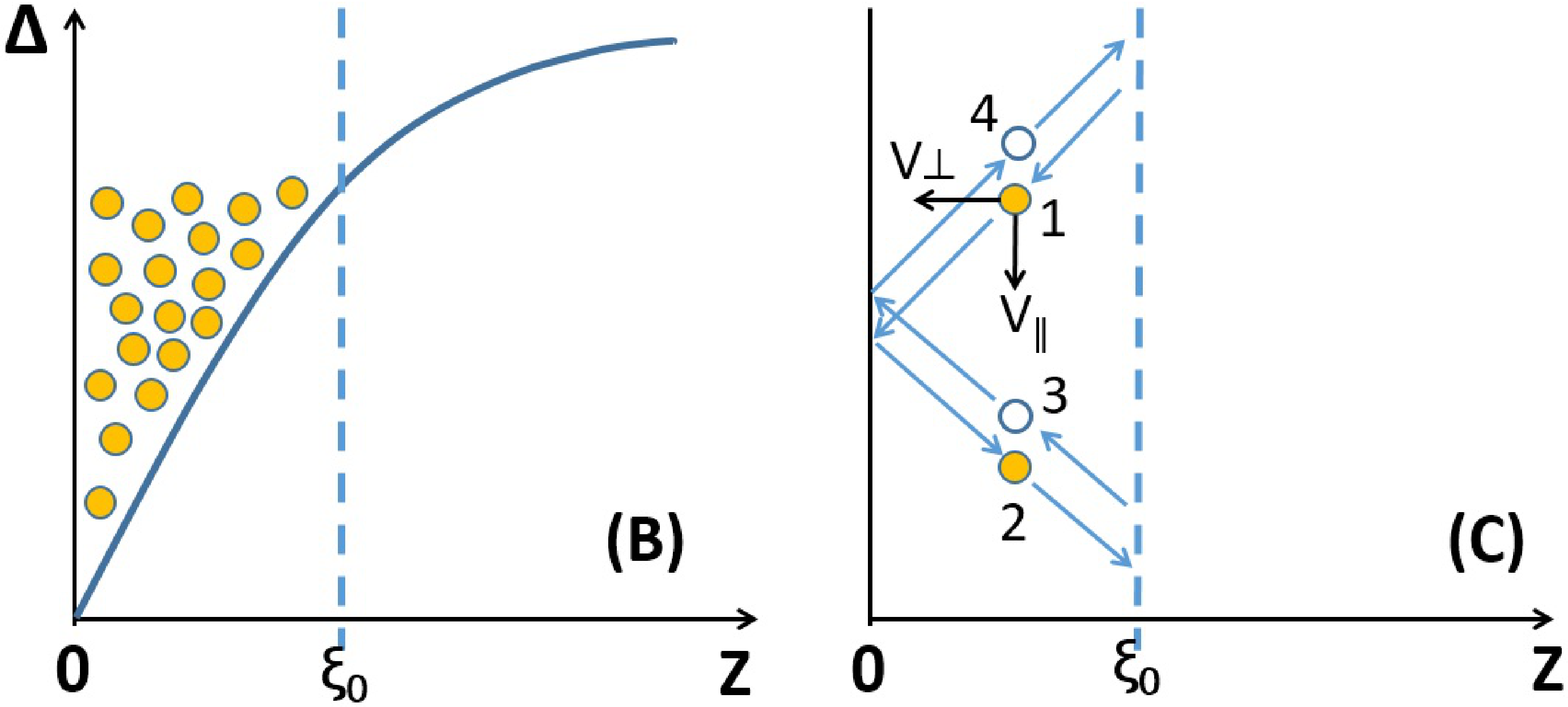}
\caption{(A) A schematic picture showing the surface density of states of superfluid $^3$He at a diffusive boundary \cite{Nagato1998JLTP1}. The quasiparticles excited in the mid-gap band are promoted by the MEMS up to the edge, $\Delta^*$. (B) The SABS confined by a potential well near the boundary at $z=0$. (C) A complete scattering cycle of a quasiparticle at a specular boundary involving two normal scatterings and two Andreev scatterings. \label{MultipleAS}}
\end{figure}

However, it is possible that the mobile plate dissipates through the surface bound states in the vicinity of the plate, leading to the additional damping. The atomic force microscopy study of the MEMS surfaces shows that the average height variation of the polysilicon surface is $\approx$ 10~nm, while their lateral size is $\approx$ 150~nm \cite{Gonzalez2013RSI1}. Since these length scales are much larger than the Fermi wavelength of the $^3$He quasiparticles, the surface of the plate is diffusive. The density of states, $D(E)$, for the surface bound states is almost independent of energy for a diffusive boundary \cite{Nagato1998JLTP1} (Fig.\,\ref{MultipleAS}). It is reasonable to project that the number of quasiparticles excited in the bound states should be proportional to temperature. Therefore, the scattering of the quasiparticles off the moving plate could lead to a linear temperature dependence of the damping, if the transverse momentum transfer occurs.

The perpendicular component of the order parameter is completely suppressed at either specular or diffusive boundary \cite{Nagato1998JLTP1}. One can expect that quasiparticles will be generated with an infinitesimally small amount of energy in the surface bound states, which are confined by the gap potential around the boundary within a distance characterized by the coherence length $\xi_0$ (Fig.\,\ref{MultipleAS}). Parallel to the plane of the plate with a specular boundary, the bound quasiparticles move with a slow velocity $v_{\|}\approx v_{_L}$, where $v_{_L}$ is the Landau critical velocity. In the direction perpendicular to the plane, however, the quasiparticle moves with a fast velocity $v_{\bot}\approx v_{_F}$, since the energy gap is closed in this direction. The Fermi velocity of the superfluid varies from 60 to 35~m/s as the pressure changes from 0 to 30~bar, while the coherence length changes from 90 to 18~nm in the pressure range. Therefore it takes approximately 1~ns for a quasiparticle to travel from the surface of the plate to the edge of the potential well, where it is then retroreflected due to the Andreev scattering. The quasiparticle becomes a quasihole and follows its previous path, moving towards the plate. It is scattered normally off the plate and Andreev scattered off the gap potential again before returning to its original position (Fig.\,\ref{MultipleAS}). This completes an entire loop involving the normal and Andreev scattering. Considering the resonance frequency of the MEMS device ($\approx 20$~kHz), one estimates that about $10^4$ such scatterings occur within one cycle of the oscillation. However, for the normal scattering at the specular boundary, there is no momentum transfer in the parallel direction between the plate and the quasiparticles, hence no damping for the shear motion of the plate. Therefore, one expects a very small damping force for the specular boundary. This may be verified by coating the MEMS plate with a couple of layers of $^4$He atoms, since the $^4$He atoms drastically alter the boundary conditions \cite{Freeman1988PRL1, Kim1993PRL1}.

For a diffusive boundary, it is difficult to trace the trajectory of a particular quasiparticle, though the process of the multiple Andreev scattering is still valid. Those having an anti-parallel group velocity component with respect to the plate velocity $v_p$ will have a higher chance of scattering, resulting in a net flux proportional to $v_p$. The tiny difference between the momentum of quasiparticles and quasiholes around the Fermi momentum accumulates due to the high number of scattering during one cycle of the plate motion. This multiple scattering process leads to a net momentum transfer between the plate and the bound states which are then promoted to higher energy states until the mid-gap edge $\Delta^*$ is reached \cite{Nagato1998JLTP1, Murakawa2009PRL1}. We propose that this process could be the underlying mechanism for the large and linear temperature dependent damping. Furthermore, our measurements in the nonlinear regime, which will be reported elsewhere, can be coherently understood with this mechanism. Nonetheless, this model neither requires the presence of a film nor involves surface bound states on the other side of the film. Currently, we do not understand the influence of another surface in close proximity on the damping of the plate. To clarify this, we have designed MEMS devices with the substrate etched away so that both sides of the plate are exposed to bulk fluid.

In conclusion, a superfluid $^3$He film with a thickness of 1.25~$\mu$m was studied by a MEMS device at various pressures. At low temperatures, an anomalously large damping on the MEMS was measured in addition to the ordinary Boltzmann damping. It was attributed to a multiple scattering picture of the interaction between the MEMS devices and the surface bound states on the film. 

% If you have acknowledgments, this puts in the proper section head.
\begin{acknowledgments}
We would like to acknowledge Peter Hirschfeld and Anton Vorontsov for helpful discussion and calculations. We also want to thank the Lancaster Low Temperature group for providing quartz tuning forks as one of the TF thermometers. This work was supported by the National Science Foundation, No.\ DMR-1205891.
\end{acknowledgments}

% Create the reference section using BibTeX:
\bibliography{Reference}

\end{document}